\documentclass[a4paper]{IEEEtran}

\usepackage[table]{xcolor}

\usepackage{rotating, subfigure}
\usepackage[final]{pdfpages}
\usepackage{url}
\usepackage{afterpage}
\usepackage{eurosym}
\usepackage{amssymb}
\usepackage{paralist}
\usepackage{longtable}
\usepackage{listings}
\usepackage[normalem]{ulem}	% For strikethrough (\sout(text})
\usepackage{bm}
\usepackage{tikz}
\usetikzlibrary{arrows}
\usepackage{amsmath}
\usepackage[noadjust]{cite}
\usepackage{tabularx}
\usepackage{balance}

\usepackage[bookmarks=false]{hyperref}

\begin{document}

%% Front Matter
%%
% Regular title as in the article class.
%
\title{The Data that Drives Cyber Insurance: A Study into the Underwriting and Claims Processes
\thanks{This paper is a preprint that has been accepted to IEEE Cyber Science 2020, International Conference on Cyber Situational Awareness, Data Analytics and Assessment (CyberSA).}}

\author{
Jason R.C. Nurse${}^{1}$, Louise Axon${}^{2}$, Arnau Erola${}^{2}$, Ioannis Agrafiotis${}^{2}$, 
Michael Goldsmith${}^{2}$, and Sadie Creese${}^{2}$\\
${}^{1}$School of Computing, University of Kent, UK\\
j.r.c.nurse@kent.ac.uk\\
${}^{2}$Department of Computer Science, University of Oxford, UK\\
\textit{\{first.last\}}@cs.ox.ac.uk
}

\maketitle

%------------------------------------------------------------------------------
% Abstract
%
\begin{abstract}
Cyber insurance is a key component in risk 
management, intended to transfer risks and support business recovery 
in the event of a cyber incident. As cyber insurance is still a new concept in practice and research,
there are many unanswered questions regarding the data and economic models that drive 
it, the coverage options and pricing of premiums, and its more
procedural policy-related aspects. This paper aims to address some of these questions 
by focusing on the key types of data which are used by cyber-insurance practitioners, particularly 
for decision-making in the insurance underwriting and claim processes. We further explore 
 practitioners' perceptions of the challenges they face in gathering and using data, and 
identify gaps where further data is required. We draw our conclusions from a 
qualitative study by conducting a focus group with a range of cyber-insurance professionals (including underwriters, actuaries, claims 
specialists, breach responders, and cyber operations specialists) and  provide valuable contributions to existing knowledge. These insights include examples 
of key data types which contribute to the calculation
of premiums and decisions on claims, the identification of challenges and gaps at various stages of data gathering, and initial perspectives on the development of a pre-competitive dataset for the cyber insurance industry. We believe an improved understanding of data gathering and usage in 
cyber insurance, and of the current challenges faced, can be invaluable for informing 
future research and practice. 
\end{abstract}

\begin{IEEEkeywords}
Cyber Insurance, Cyber risk, Underwriting, Claims, Cybersecurity, Focus Groups, Cyber Security, User Studies
\end{IEEEkeywords}

%------------------------------------------------------------------------------

\section{Introduction and background}
\label{sec:intro}
Cybersecurity incidents are now commonplace, with attackers targeting everyone from 
individuals to organisations and governments. To protect against attacks, there are a
variety of security controls, focused across the traditional areas of prevention, detection 
and reaction. These are a core part of cybersecurity risk management and can additionally support 
cyber resilience practices~\cite{nist2018,linkov2019fundamental,dupont2019cyber}. Cyber insurance
features within risk management and is a mechanism for organisations to share or 
transfer some of the risk they face. For instance, an organisation may purchase insurance 
to be covered against a data breach, and retain help recovering costs or mitigating losses 
related to it (e.g., customer notification expenses, business interruption from computer 
network downtime, incident response and costs for system restoration). As a result, cyber insurance 
has become increasingly popular, and has featured in a range of
industry and government reports/activities including those from Marsh, AXIS, 
AON, Hiscox, Deloitte, the EastWest 
Institute, ENISA and OECD~\cite{marshcb15,hiscox19,enisa16,deloitte2020oc,aon19,oecd17,ewi19,insjur18}. 

In this paper, we seek to better understand the cyber insurance process and thereby provide 
new insights into an area where there is arguably a dearth of research based on views and experience of cyber-insurance practitioners. 
Our aim is to develop an understanding of the crucial role that the types of data used in cyber insurance play in decision-making during the cyber-risk underwriting  and 
insurance claim processes. To complement this, we also look at gaps that practitioners perceive that 
exist in current data-gathering and usage processes, and explore the topic of a pre-competitive
dataset. Such a dataset could be the cornerstone for the entire cyber insurance industry in the attempt to fill gaps in data-gathering processes for underwriting and claim policies. 

Given our aim, this work involves a qualitative study, and in particular a focus group with a range 
of experienced cyber insurers from underwriting, actuarial services, claims, breach response, 
and cyber operations. Through this study, we outline a series of key data points which can 
further inform current discussions and analytics in cyber insurance, while also paving the way for future research (on which new data points may be needed to provide 
more effective and efficient insurance underwriting and claims processes). 

While cyber insurance has featured in research for at least two decades (with seminal 
works including \cite{figg2000cyber,gordon2003framework,bohme2006models,cohen2000insurance}), numerous open challenges still exist in research and practice. These span 
several issues across the field; for instance, there are the complexities of measuring, 
modelling and predicting cyber risk (especially given the evolving nature of cyber attacks, 
non-standard architectures of targets, and growing instances of `silent cyber') 
\cite{dambra2020sok,woods2019county,khalili2018designing,biener2015insurability,franke2017cyber,vakilinia2018coalitional}. 
Beyond those more technical aspects, there are also difficulties in understanding the decisions 
driving the insurance underwriting and claims processes, including those relating to
security control recommendations 
\cite{woods2017mapping,marsh19cybcat,romanosky2019content,talesh2018data}.
Furthermore, we must not overlook the reality of a lack of awareness about cyber insurance,
social insurance stigmas, and general negative perceptions 
\cite{dcms2019dbs,dcms2020dbs,hiscox19,bs18cuk,meland2017facing}. 
Such negative perceptions can inhibit the uptake of cyber insurance policies, and are boosted by the rejection of large cyber claims by 
insurance companies, which has already started to occur~\cite{nyt19}. 

In one of the most recent, comprehensive and systematic reviews of cyber insurance, 
Dambra et al.~\cite{dambra2020sok} highlight that although the field has made 
some notable strides (in areas such as game theory, economics and risk management), 
there are several open issues in risk prediction, automated data collection, catastrophe 
modelling and digital forensics. A core theme across all of these issues is data, be it for 
analytics, modelling or incident investigations. This emphasis on data can also be witnessed
in other key work, most notably in a research agenda~\cite{falco2019weis} for cyber 
risk and cyber insurance that is advocated by insurers and academics. Here, the authors call for 
research into what data should be used to assess risk and to prioritise assets, and also 
highlight the need for discussion around what cyber-related data-collection standards (that 
could drive insurance and risk analysis) would actually look like. 

Although limited, there have been efforts to clearly define the data used in 
cyber insurance. ENISA, for instance, recognising the challenges in reasoning on 
cyber risk without a shared understanding, have called for harmonisation of risk-assessment language used in cyber insurance~\cite{enisa17lang}. Within their
report, they outline a few activities and data types involved in the insurance process,
including assessing industry characteristics, audit reports and security control 
sets. Another noteworthy contribution to the field is the Cyber Exposure Data 
Schema proposed by the Cambridge Centre of Risk Studies and RSM~\cite{crs16ex}.
This provides an open resource to allow the capture, modelling and reporting of 
exposure (impacts) emerging from cyber incidents. 

Where our research differs, 
and thus adds novelty, if compared to the two aforementioned articles, is the 
focus on identifying specifically the variety of data types used within
the cyber insurance underwriting and claims processes. This work therefore
has a wider, albeit more high-level, remit than the Cyber Exposure Schema, and it
is more directed on data than ENISA's harmonisation work. Therefore, the work presented in this paper is compatible with these other articles, and may well provide
avenues for future research.

The remainder of this article is structured as follows. Section~\ref{sec:meth} 
presents the methodology we adopt to conduct our research study, and thereby
address our research aim. Next, in Section~\ref{sec:res}  we present and discuss the 
results of the study, considering the data currently used within cyber insurance 
as well as the additional data that insurers in cyber-related policies would like to have. Finally, we summarise our research and outline avenues for future work in Section~\ref{sec:concl}.

\section{Methodology}
\label{sec:meth}
To address the aims of this research, we conducted a focus group study with 
cyber insurers. Focus groups are an excellent way to explore a topic by promoting 
group discussion as applied to specific questions or problems. In our case, we were 
interested in engaging with professionals within the cyber insurance industry on the topic 
of the data that they use, or would like to have access to, in making decisions about underwriting a cyber risk and
processing an insurance claim. We prepared a series of questions targeted at this aim and grouped these into two broad categories: 
\begin{itemize}
	\item The types of data gathered at significant points in the cyber insurance process: 
	These points can include data gathered on the insurance applicant (client) before underwriting a 
	cyber risk, during the policy period, and in the event of a cyber-insurance claim. We also sought to explore the types of data that are not currently gathered but would be ideal for insurers to have. Developing 
	an understanding of these various kinds of data is an essential component in conducting research in the field 
	of cyber insurance, and creating updates or enhancements to existing platforms and solutions. 
	\item The feasibility and utility of creating a pre-competitive dataset within the cyber 
	insurance industry: This dataset could provide a shared platform for making cyber 
	insurance decisions (e.g., the value of a risk being underwritten) and thus broadly 
	advance the efforts of the entire industry, while providing a basis for ongoing cyber-insurance research. 
\end{itemize}

The focus group was designed to last 90 minutes and to be facilitated in a physical location 
 convenient to the cyber-insurance professionals who participated. We audio recorded the 
session to allow transcription at a later date, and thereby provide a richer pool of data for analysis. The 
thematic data analysis approach (\cite{braun2006using}) was adopted to allow us to assess 
the content, identify key codes (i.e., discrete information communicated in the text) and 
from these codes, construct themes (sets of related codes) based on participants' responses. 
These themes were used to extract key findings and form conclusions.

Given the aim of this study, it was imperative to recruit an experienced set of cyber-insurance professionals. We were assisted in this task by a series of research and project
contacts, who generally adhered to snowball sampling principles. This study received ethical 
approval through the university's IRB panel, all participants were informed of the purpose 
of the study and were asked to give consent prior to participation. 

\section{Results and Discussion}
\label{sec:res}

\subsection{Cyber insurance participants}
In total, a diverse group of 12 professionals from various stages of the cyber-insurance process
agreed to participate in the study. These individuals were based in the UK but engaged 
with cyber insurance portfolios both nationally and internationally, particularly in the US (which is currently the global largest market). 
Specifically, the focus group consisted of: three underwriters 
(professionals involved in assessing a cyber risk and determining whether to write a policy and 
at what premium); two actuaries (experts in measuring cyber risk and predicting financial impacts), 
two claims specialists (those who take the lead in determining whether an insurance claim arising 
from a cyber incident should be paid, and calculate the appropriate amount); two cyber training 
specialists (professionals who train client companies and cyber insurers about cyber risk 
management); a breach response specialist (personnel who focus on supporting clients in immediate 
response activities after a cyber incident has occurred); and three cyber operations experts 
(individuals specialising in various parts of operations within cyber insurance companies more 
generally). Thus, we effectively had all of the roles partaking in cyber insurance procedures represented in our focus group. It is worth noting that this group of insurers engages with various segments of
the insurance market spanning from very large to small clients and have experience with processing requests for large claims as well. This is advantageous as it provides insights 
pertaining to how insurers interact with different types of businesses and the different types of data they need for these cases. 

\subsection{Types of data gathered during the cyber insurance process}

\subsubsection{Data gathered in order to determine whether to underwrite a cyber risk}
To set the foundation for the discussions, the first question posed focused on the data currently 
gathered by insurers about potential client businesses in order to determine their level of risk, 
and thus decide whether the client should be offered a policy (or at what premium). The responses to 
this question revolved around typical organisational characteristics such as turnover, headcount, 
number of (or whether) personal records held, and sector/industry. Headcount was one of the 
most interesting data points suggested; this was because it was viewed as a way to indicate the 
size of the corporation's IT estate, and thus its potential attack surface (e.g., likelihood to be 
targeted by phishing attacks or potential for human error). 

The organisation's sector was also 
important to identify because it indicated whether certain specific regulations or procedures 
apply. For instance, a US healthcare business would need to comply with 
the Health Insurance Portability and Accountability Act (HIPAA), whereas for a manufacturing 
organisation, an insurer may be more interested in the setup of various Operational Technology 
(OT) and Information Technology (IT) systems.

Security-related information, as might be expected, also featured heavily in participants' 
responses; especially responses from underwriters. This spanned from basic data such as whether the company 
had a Chief Information Security Officer (CISO) and the extent to which employees received cybersecurity training, to more detailed 
information on IT/security setup, and the business' dependence on outsourced service providers. 
According to participants, these factors hinted at areas of significance when assessing the 
enterprise's capability (e.g., the presence of a CISO suggests the organisation may be more 
invested in cybersecurity practices), understanding of human cyber-risk (e.g., the focus on the human 
element with cyber training), and judging the potential for risks to the IT infrastructure (e.g., 
complexity of IT structure and dependence on external parties). 

There was specific mention 
of checks for security controls such as firewalls and antivirus, and the frequency with which the 
organisation updated and patched its systems. External network scans, for example those offered 
by companies such as BitSight~\cite{bitsight} and SecurityScorecard~\cite{scorecard}, were also 
used to gain an external, independent view of the security of the enterprise's infrastructure.

To complement these factors, claims specialists highlighted that loss history (analogous to works that 
capture cyber harm~\cite{harms-paper, axon2019analysing}) and the claims that these companies had made, if 
any, were also relevant factors. These would provide more insight into the organisation including 
its previous or current security practices and challenges, and their responses to them. 

In addition, participants were keen to stress that many other unquantifiable factors and data 
were important to the cyber-risk identification process. According to a cyber training specialist:

\begin{quote}
	\textit{``... and a lot of the decision-making process is unquantifiable because it's to do with 
		the interview process and the responses you're getting from the CISO who is responsible. 
		The point is that the unquantifiable stuff is at least as important as the quantifiable stuff.''}
\end{quote}

Examples of unquantifiable factors included: the experience of the cyber insurer; the way that 
the potential client answers questions posed by the insurer (e.g., their rigour and the extent 
to which they are grounded in current, as opposed to dated, technology); and the client's use 
of certain services over others (for instance, specialised legal counsel was viewed 
by insurers as more helpful after a breach than general counsel, and as such this would be a 
factor to consider depending on the client's responses). While slightly different, many of these 
aspects  can be linked to the experience of the insurer and therefore demonstrate some tacit 
knowledge that may be difficult to capture in any underwriting modelling approach (particularly 
a computational one). 

Beyond the consideration of specific data types, two other high-level themes emerged during 
our study. The first was introduced by an underwriter and centred on the reality that in addition 
to industry type, the amount of data gathered depends immensely on the size of the client's 
company. This is a key aspect as it highlights the fact that different market segments may be
approached differently by insurers. 

For smaller organisations (e.g., Small-to-Medium-sized 
Enterprises (SMEs)) for instance, it was perceived that there 
are too many possible clients, therefore it is overwhelming to collect detailed data and conduct formative assessments for each one. Furthermore, it is an extremely competitive cyber insurance market and these types of businesses will perceive such scrutiny as an obstacle for purchasing a policy. 
In such scenarios insurers may end having access to only basic information such as the proposal forms 
(e.g., \cite{woods2017mapping}) provided by the company. This is very different to larger 
organisations where there is a stronger argument for more extensive data gathering (e.g.,
meetings  with CISOs, reports and detailed presentations on the security of the organisation)
considering the higher level of the risk being underwritten. One participant from the claims 
team expressed this point clearly when speaking about smaller organisations:

\begin{quote}
	\textit{``There's a bit of a commercial trade off here on the amount of detail you go into the 
		application form because if you ask too many questions, it's too onerous, but if you 
		don't ask enough questions it comes back to bite you.''}
\end{quote}

The second theme relates to the first and pertains to the business element of the market. Participants deemed that the types of data mentioned above are useful and can help in making decisions, 
however, assessing the risk level or the security posture of a company ultimately has an element of subjectivity. 
This inherent subjectivity in defining risk may represent a potential future cost to the insurer; that is, it is a \textit{potential} 
cost not a definite one. Often, according to the participant, insurers will underwrite the risk---unless it is clearly unacceptable---and put several protective measures around it to 
secure the business. This suggests that in some cases, the need to increase business 
(e.g., clients, revenue, etc.) can outweigh the need to make the `perfect' underwriting decision. 
This fact is important especially as it focuses on the reality of market forces instead of a drive 
for perfection.

We also provided to participants the opportunity to think about and propose data that is not currently 
gathered, but which might be helpful to them when making their decisions on the cyber security 
posture of an organisation. 

Overall, participants' responses to this question concentrated mainly 
on the security features of the organisation and on obtaining further insight into such processes. In 
particular, there was a desire to know more about the training and awareness measures 
undertaken to protect against threats (and human errors), and the extent to which backups 
were created, maintained and tested. These factors suggested a primary focus on the human 
side of risk (given that it was viewed as a gateway to numerous current attacks) and the ability of 
businesses to recover from incidents (with suitable backups). 

One participant, a breach 
response expert, also suggested that external assessments (e.g., reports from penetration 
testing companies) would be ideal to have access to, as they provided an independent security review of 
the organisation. While these might not be trusted fully (as they was not commissioned by the 
insurers themselves), it could provide additional input to the decision process. 

Another aspect mentioned pertained to understanding more about the business' plans over
the upcoming year. For instance, plans to migrate IT systems to the cloud, to update 
IT systems, to change primary firewalls, or to acquire other organisations; if such plans exist, insurers were interested in what protective  
activities are introduced in these cases. One underwriter commented:

\begin{quote}
	\textit{``We're just told that [something may happen] and no information about what's 
		actually going on; let's say it's a migration to the cloud or M\&A [Mergers and Acquisitions] 
		activity and they are interacting with another company, we need to know what that 
		process is, what the road map is, any rollback contingency plans in place.''}
\end{quote}

This suggests that insurers currently may receive general information, but not at the level 
where they can adequately understand the risk involved. This is an interesting point 
considering the rate at which an underwritten cyber risk could change depending on the 
specifics of any of the aforementioned changes (e.g., a migration of systems to the cloud). 

In discussing the topic of IT systems, one actuary suggested that it would be ideal to be 
able to have more insight into organisational processes and dependencies, in order to allow 
insurers to better consider risk aggregation in decision making. Specifically:

\begin{quote}
	\textit{``If you're an insurance company and you've written a thousand policies, the key 
		issue for us is getting what services they are using, what providers, do they all have 
		Amazon Web Services, and how reliant are they on it.''}
\end{quote}

This perspective focuses on the underlying requirement for insurers to understand 
more about how client organisations work. This is not only for defining value at risk and premiums,
but also to elucidate systemic risk across their client insurance portfolio. Systemic risk is a 
crucial concern for insurers and has been explored in detail in various reports~\cite{eurosrb20,ewi19,rand19}.

There were also calls for more information due to ambiguities in proposal forms returned 
by potential clients. The wording of these forms was, at times, viewed as too rigid and 
not flexible enough to cater for large numbers of organisations. This would be particularly
important during engagement with smaller organisations, where proposal forms are a 
primary method of data gathering to make decisions on risk exposure. For instance, a question 
may ask, ``Do you perform penetration tests on a quarterly basis? Yes / No''. This can, 
on occasion, force a company to select ‘No' even if they perform tests more regularly; this 
can therefore be contrary to what the question is aiming to assess. 

Moreover, according 
to one underwriter, standardised forms from some brokers may focus on topics not 
relevant to all clients. An example was provided that suggests that most of the forms currently 
concentrate on privacy (likely largely due to related laws and regulations such as the General Data Protection Regulation~\cite{gdpr}) 
and business interruption. However, this emphasis may be unsuitable for some 
organisations, e.g., a proposal form with many privacy questions is arguably not best 
suited for a manufacturer. 

The last point raised offered a different opinion and suggested that while having a good understanding about security controls was useful, it would be better to know about the effectiveness 
of the company's controls at addressing the risk they face. Summing up the point, a cyber 
operations specialist said:

\begin{quote}
	\textit{``For me, a lot of this is around effectiveness. All these things we are talking 
		about, are they effective, so is training effective? If you have a training programme in 
		place and it's busy, it's automated and yet it isn't working and in fact staff are failing 
		more phishing tests. So, it might tick a box on the proposal form that says, yep we 
		got an automated cyber awareness training, brilliant, but it doesn't then say that 
		everyone's failing it and failing it more and more each time.''}
\end{quote}

This is a salient point as it highlights the fact that security is more than the presence of 
controls, it must also consider their effectiveness. It further raises the question of how effectiveness can best be assessed; should companies be required to capture and 
present this information, or should effectiveness be part of a larger framework, such as one 
that would compose a pre-competitive cyber insurance dataset? In the former case, one 
participant expressed that it might be difficult to achieve this as it could, in effect, `show 
up' CISOs/IT managers that are not performing well. Given that they are often the ones with 
whom insurance brokers and underwriters interact, such company representatives would be increasingly reluctant to share security-control effectiveness information. Finally, there is the reality that 
while  controls, tools and training may demonstrate increased effectiveness (e.g., more 
intrusions blocked or reduced phishing click rates), attackers only need to be successful 
once to compromise systems~\cite{nursecybercrime}. 

\subsubsection{Data gathered between writing a policy and its renewal}
The next area we explored considered the data gathered by insurers on their clients in the 
period between writing the policy and its renewal (typically 12 months later). In general, 
participants mentioned that only minimal amounts of data was gathered at this stage, if any. 
The most common information of interest to underwriters and claims specialists was whether 
there were material changes to the client's business, and in particular whether the client 
organisation acquired any other companies or whether they were acquired by others. 
While this tended to be the standard way of monitoring changes to clients in the insurance 
industry, this was also an area of concern because of the thresholds set before clients 
were obliged to notify insurers. As one cyber underwriter stated:

\begin{quote}
	\textit{``It depends on the broker, the only ones [causes for insurer notification] 
		that are written into the policy are changes in the controls if they are acquired or 
		acquire someone else but you normally have a threshold for that, so 15\% or so 
		of your revenues and if it's under that then they're not obliged to tell us during 
		the policy period.''
}
\end{quote}

This suggests that information on risk may be lacking during this period, as one might 
imagine a situation where, for example, two million-dollar organisations merge, and due 
to their similar sizes, the 15\% revenue difference threshold is not met. As such, there may 
be no updates provided to the insurer. According to participants, there are also no 
contractual requirements for such a notification/update during the policy (unless clauses 
have been specific, e.g., pertaining to thresholds). Acquisitions can also be important to 
examine from the perspective of the risk profiles of organisations. As described by one 
underwriter:

\begin{quote}
	\textit{``... say a university acquires a payment processor which would change the risk 
		profile quite dramatically but [as] they [i.e., the payment processor] are tiny, you might 
		not even know because it wouldn't trigger that acquisition threshold in the policy.''
}
\end{quote}

This raises a crucial point linked to how dynamically and significantly risk profiles of organisations 
can change during the lifetime of a policy, without the knowledge of the insurers who have 
underwritten that risk. When questioned about whether there was an opportunity to 
gather more data on client operations in the process leading up to a policy renewal, 
participants mentioned that it was possible but also challenging. One underwriter noted 
that it was not uncommon to receive policy renewal applications, via a broker, from large 
client organisations that only checked whether the organisation had changed their 
business or had any claims in the last 12 months. This hinted at the challenges of working 
with insurance brokers, who can often act as the gatekeepers and primary interface to some clients, 
and to the industry more fundamentally considering the fact that if not reviewed, cyber 
risk can change significantly over such long periods. An underwriter picked up on exactly 
this point with the comment:

\begin{quote}
	\textit{``The thing about cyber is that if a company hasn't made any changes or improvements 
		in 12 months, that should lead to a premium increase because the risk is very different.''
}
\end{quote}

However, and as highlighted earlier, the business component often mediates such 
decisions, with another underwriter quickly interjecting and stating that the insurer would 
never be able to implement or follow through with that. The first underwriter then continued:

\begin{quote}
	\textit{``But then, there's the side from a business or a competitive market standpoint: 
		if a company comes in and says: no changes for the last 12 months. Then, we put the 
		premium up 5 or 10\% because they haven't made any positive changes, then we will 
		lose the business because someone will just underwrite it at the price that we did last 
		year. So, there is that side of it as well.''
}
\end{quote}

The difficulty in such cases therefore resides not only in identifying and gathering 
appropriate data about a client and their potential value at risk, but also in balancing 
this with the need to remain competitive in the market. This tension is an intriguing one 
noting how quickly risks can change. Ultimately, it may also mean that actual client risk 
profiles may not accurately align to how that risk is viewed on an insurer's books (and 
also reflected in the premiums charged).   

Similar to our previous section, we allowed participants to suggest data that is not 
currently gathered during this period, but which they regarded as of interest to them. 
This resulted in a largely homogeneous set of responses, including progress on planned 
system or process migrations (e.g., system updates, IT changes, migrations of data to 
other platforms), and updates on security training activities. Some participants were keen 
to discover more about the security maturity of the organisation and how they 
responded to security developments (including incidents). To sum it up, a breach 
response specialist commented:

\begin{quote}
	\textit{``More information on the wins and losses and how they handle them; and 
		how they get systems back up and running and how effective are they at doing 
		that most of the time.''
}
\end{quote}

This is noteworthy as it suggests an appetite --- at least for some part of the insurer 
community --- for data on effectiveness of security mechanisms and processes. The 
benefit in such cases is that they appreciate the reality of data breaches and therefore 
concentrate on response capabilities as well.

\subsubsection{Data gathered after a cyber incident}
To further expand our understanding of data gathered by insurers, we then moved to 
consider what data was gathered after an incident. It was clear from responses that in 
the event of a claim, insurers were able to gather a significant amount of data about the 
organisation. This could span follow-up questions in line with the proposal forms (that 
were gathered before underwriting the risk) to compare those statements with actual 
activities, changes in systems, who has been contacted/involved in dealing with the 
incident, what exactly has occurred and any associated costs. These were aptly 
summarised by one cyber insurance claims specialist:

\begin{quote}
	\textit{``First and foremost, the two prime responsibilities are to try to find out as 
		early as possible what's the claim going to cost so that we can reserve and to find 
		out enough information to confirm or deny coverage as early as possible.''}
\end{quote}

Another source of valuable data was the breach response team involved in handling the 
incident. These teams would provide a detailed capture of how the business is actually 
operating and specific insight behind the cause and chain of activities involved in the cyber 
security incident. 

While data access after an incident was not a challenge, participants highlighted that data 
gathering and storage was. Typically claims would need to be first stored in a standard 
market platform, and then this platform accessed to retrieve 
information about the incident. The difficulty with the platform however, was that it 
significantly constrained what data could be uploaded about 
an incident. For instance, participants identified that it was not possible to easily note the 
types of attack or specifics of malware involved. This 
would instead need to be shared over some other platform such as email. These workarounds 
were viewed as less than ideal and had further implications on the ability to search for past 
incidents, and run analytics on risks (e.g., understanding what risks clients were most 
exposed to within the last 12 months). This is an area that would require further work, 
most likely at the industry level.

\subsection{The feasibility and utility of creating a pre-competitive dataset
}
Having gained insight into the various types of data that is gathered at key points in 
the cyber-insurance process, we shifted our attention to investigating the feasibility and 
utility of the creation of a pre-competitive dataset within the cyber-insurance industry. 
The goal of this dataset would be to provide a platform for making cyber-insurance decisions 
that could be shared across industry.

When posed the question about the feasibility of the creation of such a dataset, 
none of the participants felt that it would work, nor were they comfortable in sharing 
the information necessary to create it. The justification for this decision was cited to 
be building and maintaining a competitive advantage, in what is still a new market. As 
one underwriter made clear:

\begin{quote}
	\textit{``[As a cyber insurer, you've] taken the risk to build to where you are, that's your 
		IP at that point, that's your competitive advantage.''
 }
\end{quote}

And as emphasised by an cyber-insurance actuary:

\begin{quote}
	\textit{``It's general economics, the first adopter or first mover advantage, given you've 
		invested heavily into being market leaders, why should you want to enable other people 
		to come in and compete with you essentially?'' 
 }
\end{quote}

In discussing the feasibility of creating such a dataset further, another participant suggested that the structure of a 
pre-competitive dataset may already exist in the form of the proposal forms (such as 
those summarised in research~\cite{woods2017mapping}) issued by insurers. These 
forms gather relevant data about clients and can be shared across underwriters, and 
therefore, if the client information from each completed form was collected and placed 
in a database, that could form its basis. This was an intriguing suggestion, 
but one that was quickly opposed by another participant, an actuary, who highlighted 
the fact that companies and even clients would not be willing to share this data. He noted:

\begin{quote}
	\textit{``I would argue that proposal forms are also proprietary and you wouldn't 
		want to share that with anyone ... Some of our clients don't even want to send 
		us their information as their broker, let alone share it with their re-insurers. 
		People are insanely protective.''
 }
\end{quote}

These comments also relate to earlier findings and the difficulty in gathering data from 
clients. In this case, the challenge was not only gathering that data but 
encouraging insurers to share it (or/and other information) into a collaborative 
pre-competitive dataset. 

From these and other comments made during the focus group, it was apparent that,  
at least based on this group, the market may not be ready to create a pre-competitive 
dataset. This is linked to the fact that insurers may have invested significantly in becoming 
market leaders and that the introduction of such a dataset would negatively impact their efforts (i.e., 
either by lowering barriers to enter the market or by exposing current market knowledge). 

Although the idea of creating a pre-competitive dataset may not be currently feasible, 
we were interested to gather participants' opinion on what types of data would be ideal to 
include in such a dataset. We believe this would be useful information to discuss in this 
setting considering that the market's opinion on such a dataset might change as time 
progresses. This question resulted in a few different responses. One cyber training 
specialist expressed that client sector, turnover, number of employees and number of 
customers were enough to determine 80\% of the answer to how to price risk. In his 
opinion, this, in addition to whether the client had a claim or not in the past, could lead 
to a reasonable judgement. As such, these data points would be crucial to a 
pre-competitive dataset. 

A claims specialist offered a different perspective and suggested that the more 
confidential information (e.g., past incident information, actual cost of claims and 
breakdown of costs, etc.) would be more useful at making decisions at that stage.
It may therefore be possible that different levels of data (in a pre-competitive 
dataset) are required at separate stages of the process. This also, of course, needs 
to consider appropriate laws, regulations and client preferences.

We further sought to explore participants' opinions on the extent to which data on assets, 
threats, harms and controls may feature as a part of the dataset. According to one actuary 
however, the cyber insurance industry was not at that stage of maturity as yet. In summary:

\begin{quote}
	\textit{``That information may be too hard to capture because every organisation will have 
		different types of assets, different security setups and to standardise that and draw 
		meaningful conclusions from it --- we're not at that stage yet.''
 }
\end{quote}

While an isolated response, this does highlight a real challenge behind gathering such 
a dataset and standardising it to the point that analytics can be performed by sector, 
size or other enterprise characteristics. Focusing specifically on security controls, we 
also sought to explore which controls participants viewed as the most effective, and 
would, for instance positively impact their decision of taking on a particular risk. There 
were various responses to this question including cybersecurity training, regular 
penetration testing, network segmentation, multifactor authentication, dual verification 
of payments, and data monitoring and control. The most common response however was 
in the organisation's ability to respond to a cyber incident. One actuary commented: 

\begin{quote}
	\textit{``It's not if it's when, it's how you handle post breach. Have you got a PR 
		statement prepared? How do you minimise the damage?''
 }
\end{quote}

To build on this, participants stressed the importance of practising incident 
response and rehearsing how to respond in such cyber incident situations. This included 
identifying appropriate communication messages and settings.
It was clear that members of the focus group were aware of the pervasiveness 
of attackers and the high likelihood of eventual breaches. These are all common 
principles that can be found in most cyber incident response or resilience playbooks.

\section{Conclusions and Future work}
\label{sec:concl}

Cyber insurance is still a field in its infancy, and as such, there are several open 
questions pertaining to assessing cyber risk, encouraging cyber insurance adoption, 
calculating risk exposure, writing policies, and 
supporting claims and business recovery. This paper has contributed to the field by providing 
new insight into the types of data which cyber-insurance practitioners use on a daily basis to conduct 
their business. While our work has primarily engaged with UK-based participants,
their experience with global portfolios means that our findings are relevant for all geographical areas. 
This is important given how quickly the cyber insurance market is expanding worldwide. 

From our analysis of the focus group data, we identified a large range of data types gathered by 
insurers, and reported these within the main stages in which they are used. 
For instance, before a risk is underwritten, insurers are likely to be interested in security 
related information such as whether the company has a CISO and the extent to which 
employees received cybersecurity training. At the claim stage, the amount of information 
gathered can drastically increase and is often an opportunity to clarify key assertions made 
earlier in the process.

Throughout this data exploration exercise, it became apparent that generally cyber 
underwriters have a challenging task balancing the gathering of data from clients. If 
too much data is requested clients (or potential clients) may choose a competitor but 
if too little data is requested, it may increase the risk to the insurer. Claims specialists 
also have challenges to overcome even though after an incident they do receive a significant 
amount of data. A primary issue here is to have in place the systems meant to capture data, and to design the platforms where data can be easily searched and analysed. 

On the topic of the 
creation of a pre-competitive dataset, participants did not view this favourably. Their 
perspective was motivated by the impact of such a dataset on the competitive advantage. 
There was also the question of exactly what such a dataset would contain. While some 
individuals provided suggestions, these did not always align and such conflicts clearly represent the 
challenge of creating a pre-competitive dataset at present. 

This work provides directions for several avenues of future work. The first involves expanding upon this 
research with a large-scale survey with cyber-insurance practitioners. Focus groups provide a perfect 
opportunity to explore topics in detail, however surveys allow such insights to be 
expanded upon and, to some extent, generalised. 
In particular, it would be valuable to use surveys to further explore the extent of the 
identified challenges to, and gaps in, current data collection and the reasons 
behind them. This could inform potential solutions that align with the 
capacity and requirements of the insurance community. We could also aim to examine
in more detail the various market segments that exist, and how data use and needs
by insurers may vary across these segments. It was clear from our work that larger organisations
are subject to more exhaustive data requests however, we are yet to explore what 
specific types of data may be preferred by insurers depending on a company's market segment. 

A second area of research could 
focus on operationalising the data points mentioned (and any 
other data points that can be discovered), through the definition of a comprehensive model or 
end-to-end cyber-insurance process. This could specify key inputs and outputs, which 
can impact risk exposure and premiums. Such models or processes can be immensely 
valuable for research (i.e., in providing insights where academic and research efforts 
may be concentrated), but may be against the preferences of cyber insurers --- this 
links to the resistance to the creation of a pre-competitive dataset.

\section* {Acknowledgements}
This research was sponsored by AXIS Insurance Company, whose support is 
gratefully acknowledged.

%------------------------------------------------------------------------------
%
\balance
\label{sect:bib}
\bibliographystyle{abbrv}
\bibliography{refs}
\end{document}